\documentclass[a4paper]{spie}  

\usepackage[]{graphicx}
\usepackage{amssymb,amsmath,psfrag,url}

\newcommand{\Field}[1]{{\boldsymbol{#1}}}

\newcommand{\Kvec}[1]{{\vec{#1}}}

\title{Numerical Investigation of Light Scattering\\ off Split-Ring Resonators} 

\author{
Sven Burger\supit{\,a}, 
Lin Zschiedrich\supit{\,a},
Roland Klose\supit{\,a}, 
Achim Sch{\"a}dle\supit{\,a},
Frank Schmidt\supit{\,a}, 
Christian~Enkrich\supit{\,b}, 
Stefan Linden\supit{\,b},  
Martin Wegener\supit{\,b}, and
Costas M. Soukoulis\supit{\,c}
\skiplinehalf
\supit{a}
Zuse Institute Berlin,
Takustra{\ss}e 7,
D\,--\,14\,195 Berlin,
Germany\\
DFG Forschungszentrum {\sc Matheon},
Stra{\ss}e des 17.~Juni 136, 
D\,--\,10\,623 Berlin,
Germany\\
JCMwave GmbH,
Haarer Stra{\ss}e 14a,
D\,--\,85\,640 Putzbrunn, 
Germany
\skiplinehalf
\supit{b}
Institut f. Nanotechnologie, Forschungszentrum Karlsruhe,
Institut f. Angewandte Physik, Universit\"at Karlsruhe (TH),
DFG Forschungszentrum f. Funktionelle Nanostrukturen (CFN),
Wolfgang-Gaede-Str. 1,
D\,--\,76\,131 Karlsruhe,
Germany
\skiplinehalf
\supit{c}
Ames Labs and Dep. of Physics and Astronomy,
ISU, Ames, Iowa 40\,011, USA\\
Foundation for Research and Technology (FORTH), 71\,110 Heraklion, Crete, Greece
}

\authorinfo{
Further author information: (Send correspondence to S.B.)\\
URL: http://www.zib.de/nano-optics/\\
S.B.: E-mail: burger@zib.de, Telephone: +49 30 84185 302\\
-- \\
Proc.~SPIE {\bf 5955}, p.~18\,-\,26 (2005).
}

\begin{document} 
  \maketitle 

\noindent
Copyright 2005  Society of Photo-Optical Instrumentation Engineers.\\
This paper was published in Proc.~SPIE {\bf 5955}, p.~18\,-\,26 (2005), 
({\it Metamaterials}\,; 
Tomasz Szoplik, Ekmel \"Ozbay, Costas M. Soukoulis, Nikolay I. Zheludev; Eds.).
and is made available 
as an electronic reprint with permission of SPIE. 
One print or electronic copy may be made for personal use only. 
Systematic or multiple reproduction, distribution to multiple 
locations via electronic or other means, duplication of any 
material in this paper for a fee or for commercial purposes, 
or modification of the content of the paper are prohibited.

\begin{abstract}
It seems to be feasible in the near future to exploit the 
properties of left-handed metamaterials in the telecom or 
even in the optical regime.
Recently, split ring-resonators (SRR's) have been realized 
experimentally in the near infrared (NIR) and optical regime~\cite{Linden2004a,Enkrich2005a}.
In this contribution we numerically investigate light 
propagation through an array of metallic SRR's in the 
NIR and optical regime and compare our results to experimental 
results.

We find numerical solutions to the time-harmonic Maxwell's 
equations by using advanced finite-element-methods (FEM). 
The geometry of the problem is discretized with unstructured 
tetrahedral meshes. 
Higher order, vectorial elements (edge elements) are used as 
ansatz functions. 
Transparent boundary conditions 
(a modified PML method~\cite{Zschiedrich2005b}) 
and periodic boundary conditions~\cite{Burger2005a} are implemented, 
which allow to treat light scattering problems off 
periodic structures. 

This simulation tool enables us to obtain transmission and 
reflection spectra of plane waves which are incident onto the 
SRR array under arbitrary angles of incidence, with arbitrary 
polarization, and with arbitrary wavelength-dependencies of 
the  permittivity tensor. 
We compare the computed spectra to experimental results and 
investigate resonances of the system.

\end{abstract}


\section{Introduction}

With the advances in nanostructure physics it has become possible 
to manipulate light on a lengthscale smaller than optical 
wavelengths~\cite{Linden2004a,Enkrich2005a}.
It is now possible to construct periodic structures 
made of artifical nanostructures with macroscopic 
properties which do not occur in nature.
These nanostructures are large on the atomic scale, therefore they 
can be of complex geometry. But they are small on the scale of the 
wavelength of the illuminating light, 
therefore the system has properties of an effective homogeneous 
material, in particular a macroscopical electric permittivity, 
$\varepsilon$, and magnetic permeability, $\mu$.

Of special interest are metamaterials with both, negative $\varepsilon$
and negative $\mu$~\cite{Veselago1968a}. 
This leads to a negative index of refraction~\cite{Smith2000a,Shelby2001a} and
allows in principle to overcome limits in the resolution of 
optical imaging systems~\cite{Pendry2000a}.
 
\section{Arrays of Split-Ring Resonators}
Split-ring resonators can be understood as small $LC$ circuits 
consisting of an inductance $L$ and a capacitance $C$.
The circuit can be driven by applying external electromagnetic 
fields.
Near the resonance frequency of the $LC-$oscillator
the induced current can lead to a magnetic field opposing the external 
magnetic field. 
When the SRR's are small enough and closely packed -- such that the system 
can be described as an effective medium -- 
the induced opposing magnetic field corresponds to an effective 
negative permeability, $\mu<0$, of the medium.

\begin{figure}[hbt]
\centering
\psfrag{(a)}{(a)}
\psfrag{(b)}{(b)}
\includegraphics[width=\textwidth]{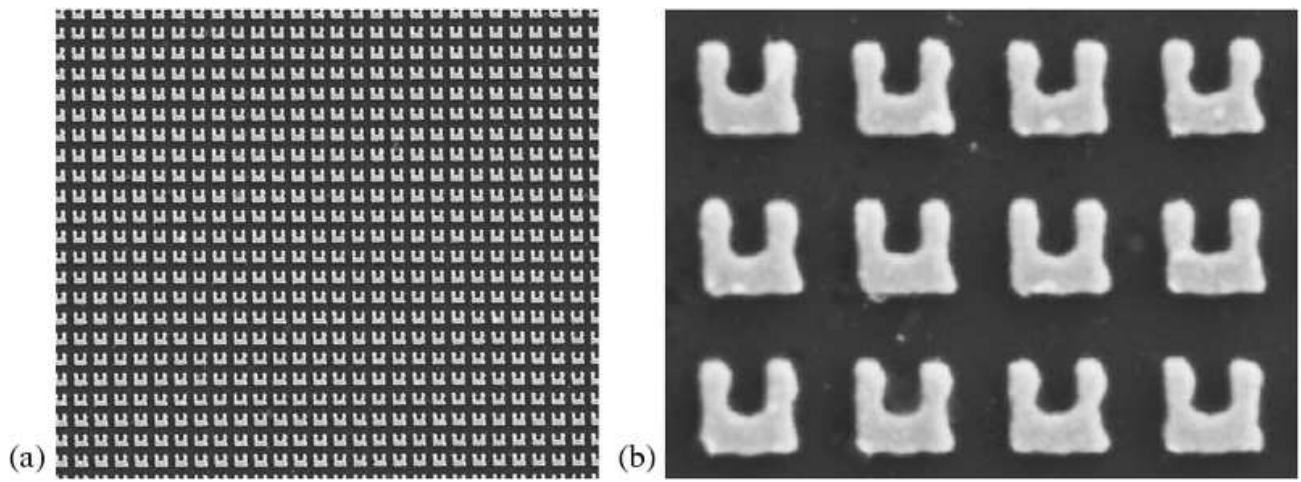}
\caption{
Electron micrographs of an array of SRR's with different magnifications.
Field of view corresponds to approximately 
$8.5\,\mu$m$\times 7.6\,\mu$m (a), 
resp.~$1.25\,\mu$m$\times 1.0\,\mu$m (b). 
The total size of the fabricated array is $100\,\mu$m$\times 100\,\mu$m.
{\footnotesize (See original publication for images with higher resolution.)}
}
\label{micrograph}
\end{figure}

Arrays of SRR's with resonances in the NIR and in the optical 
regime have been fabricated using electron-beam lithography 
on a 1\,mm thick glass substrate coated with a 5\,nm thick 
film of indium-tin-oxide (ITO).
Figure~\ref{micrograph} shows electron micrographs of a produced 
sample.  
The 'U'-shaped SRR's are produced with a thickness of $\Delta z=30\,$nm.
Details on the production can be found in previous 
works~\cite{Linden2004a,Enkrich2005a}.

Due to the small dimensions of the $LC$ circuits their resonances are 
in the NIR and optical regime~\cite{Enkrich2005a}.
The elementary vectors of the periodic array, 
$\vec{a}_1=(315\,\mbox{nm},\,0,\,0)$ and
$\vec{a}_2=(0,\,330\,\mbox{nm},\,0)$,
are smaller than NIR and optical wavelengths, therefore in these 
regimes the system is well described as an effective medium.

In what follows we investigate numerically the transmission of light 
under oblique incidence through these samples.

\begin{figure}[hbt]
\centering
\psfrag{}{}
\psfrag{(a)}{}
\psfrag{ax}{$a_x$}
\psfrag{ay}{$a_y$}
\psfrag{lx}{$l_x$}
\psfrag{ly}{$l_y$}
\psfrag{t}{$\Delta z$}
\psfrag{d}{d}
\psfrag{w}{w}
\psfrag{x}{x}
\psfrag{y}{y}
\psfrag{z}{z}
\psfrag{gold}{gold}
\psfrag{ITO}{ITO}
\includegraphics[width=0.45\textwidth]{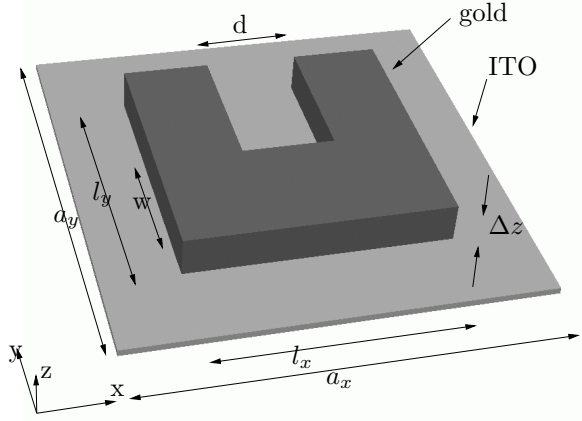}
\caption{Geometry of the unit cell in the array of SRRs. 
         The surrounding air and the substrate are not indicated.
{\footnotesize (See original publication for images with higher resolution.)}
}
\label{geometry}
\end{figure}

\section{Light Propagation in Periodic Arrays of SRRs}
We consider light scattering off a system which is 
periodic in the $x-$ and $y-$directions and is enclosed by  homogeneous 
substrate (at $z_{sub}$) and superstrate (at $z_{sup}$) 
which are infinite in the $-$, resp.~$+z-$direction. 
Light propagation in this system is governed by Maxwell's equations
where we assume vanishing densities of free charges and currents.
The dielectric coefficient $\varepsilon(\vec{x})$ and the permeability 
$\mu(\vec{x})$ are periodic and complex, 
$\varepsilon \left(\vec{x}\right)  =  \varepsilon \left(\vec{x}+\vec{a} \right)$, 
$\mu \left(\vec{x} \right)  =  \mu \left(\vec{x}+\vec{a} \right)$.
Here $\vec{a}$ is any elementary vector of the periodic lattice~\cite{Sakoda2001a}.  
For given primitive lattice vectors 
$\vec{a}_{1}$ and $\vec{a}_{2}$ the elementary cell 
$\Omega\subset\mathbb R^{3}$ is defined as
$\Omega = \left\{\vec{x} \in \mathbb R^{2}\,|\,
x=\alpha_{1}\vec{a}_1+\alpha_{2}\vec{a}_2;
0\leq\alpha_{1},\alpha_{2}<1
\right\}
\times [z_{sub},z_{sup}].$
A time-harmonic ansatz with frequency $\omega$ and magnetic field 
$\Field{H}(\vec{x},t)=e^{-i\omega t}\Field{H}(\vec{x})$ leads to
the following equations for $\Field{H}(\vec{x})$:
\begin{itemize}
\item
The wave equation for the magnetic field:
\begin{equation}
\label{waveequationH}
\nabla\times\frac{1}{\varepsilon(\vec{x})}\,\nabla\times\Field{H}(\vec{x})
- \omega^2 \mu(\vec{x})\Field{H}(\vec{x}) = 0,
\qquad\vec{x}\in\Omega,
\end{equation}
\item
The divergence condition for the magnetic field:
\begin{equation}
\label{divconditionH}
\nabla\cdot\mu(\vec{x})\Field{H}(\vec{x}) = 0,
\qquad\vec{x}\in\Omega,
\end{equation}
\item
Transparent boundary conditions at the boundaries to the 
substrate (at $z_{sub}$) and superstrate (at $z_{sup}$), $\partial\Omega$,
where $\Field{H}^{in}$ is the incident magnetic field (plane wave 
in this case), and $\vec{n}$ is the normal vector on $\partial\Omega$:
\begin{equation}
\label{tbcH}
	\left(
        \frac{1}{\varepsilon(\vec{x})}\nabla \times (\Field{H} - 
        \Field{H}^{in})
	\right)
	\times \vec{n} = DtN(\Field{H} - 
        \Field{H}^{in}), \qquad \vec{x}\in \partial\Omega.
\end{equation}
The $DtN$ operator (Dirichlet-to-Neumann) is in our case realized with 
the PML method (perfectly matched layer)~\cite{Zschiedrich2005b}. 
It is a generalized formulation of Sommerfeld's radiation condition and
can be realized alternatively, e.g., by the Pole condition method~\cite{Hohage03a,Hohage03b}.
\item
Furthermore, 
the Bloch theorem applies for wave propagation in periodic media.
Therefore we aim to find Bloch-type field distributions~\cite{Sakoda2001a} 
solving
Equation~(\ref{waveequationH}), 
defined as
\begin{equation}
\label{bloch}
\Field{H}(\vec{x}) = e^{i \Kvec{k}\cdot\vec{x}} \Field{u}(\vec{x}), \qquad
\Field{u}(\vec{x})=\Field{u}(\vec{x}+\vec{a}),
\end{equation}
where $\vec{a}$ is any elementary vector of the periodic lattice
and the Bloch wavevector $\Kvec{k}\in\mathbb{R}^3$ is defined by the
incoming plane wave $\Field{H}^{in}$.
\end{itemize}

Similar equations are found for the electric field 
$\Field{E}(\vec{x},t)=e^{-i\omega t}\Field{E}(\vec{x})$.

For applying the method of finite elements to solve
Equations~(\ref{waveequationH}) -- (\ref{bloch}), resp.~the
corresponding equations for the electric field, it is necessary to 
reformulate these in the weak form. 
Care has to be taken in the right choice of functional spaces
which are then discretized by Nedelec's edge elements. 
This leads to  a large sparse matrix equation (algebraic problem).
For details on the weak formulation, 
the choice of Bloch-periodic functional spaces,
the FEM discretization, and our implementation of the PML
method we refer to previous works~\cite{Zschiedrich2005b,Burger2005a,Zschiedrich2005a}.

\begin{table}[h]
\centering
\begin{tabular}{|rr|}
\hline
$a_x$ [nm] & 315.0\\
$a_y$ [nm] & 330.0\\
$l_x$ [nm] & 200.0\\
$l_y$ [nm] & 200.0\\
$w$ [nm] & 90.0\\
$d$ [nm] & 70.0\\
$\Delta z$ [nm] & 30.0\\
$\Delta z_{ITO}$ [nm] & 5.0\\
\hline
$\epsilon_{Substrate}$ & 2.25\\
$\epsilon_{ITO}$&3.8\\
$\epsilon_{air}$&1.0\\
$\epsilon_{gold}$&Drude model ($\omega_p, \omega_c$)\\
$\omega_p$ [s$^{-1}$]&$1.367\cdot 10^{16}$\\
$\omega_c$ [s$^{-1}$]&$6.478\cdot 10^{13}$\\
\hline
\end{tabular}
\caption{Geometrical parameters and material parameters for the 
SRR simulations corresponding to the experimental 
measurements reported by Enkrich et al~\cite{Enkrich2005a}.
{\footnotesize (See original publication for images with higher resolution.)}
}
\label{srr_para_t5}
\end{table}

To solve the algebraic problem on a standard personal computer we use 
either standard linear algebra decomposition techniques (LU-factorization)
or iterative methods, 
depending on the problem size.
Multi-grid algorithms~\cite{Deuflhard2003a} are used within the algorithm 
for preconditioning. 
As finite-element ansatz functions, we typically choose edge elements 
of second order~\cite{Monk2003a}.
Due to the use of multi-grid algorithms, the computational time and the memory requirements
grow linearly with the number of unknowns.  

\section{Spatial discretization and material parameters}
\label{parameterchapter}
\begin{figure}[hbt]
\centering
\includegraphics[width=0.45\textwidth]{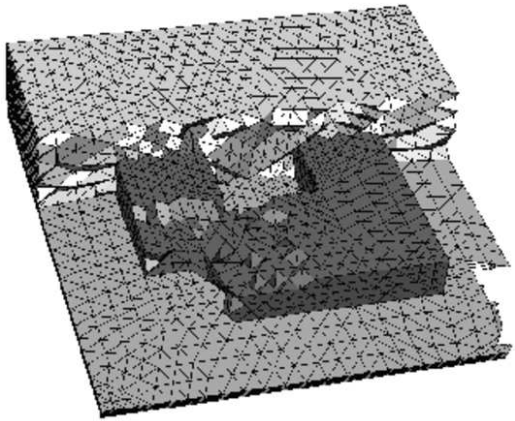}
\caption{Visualization of a part of the tetrahedra of the spatial discretization 
of the SRR geometry shown in Figure~\ref{geometry}. Dark gray tetrahedra: gold SRR; light 
gray: air; gray: ITO. Prism elements discretizing the exterior domain 
are not shown.{\footnotesize (See original publication for images with higher resolution.)}
}
\label{mesh}
\end{figure}

Figure~\ref{geometry} shows the geometry of the unit cell of the array of 
investigated split-ring resonators. 
The geometrical parameters (corresponding to the experimentally 
realized SRR, see Fig.~\ref{micrograph}) are listed in Table~\ref{srr_para_t5}.
This unit cell corresponds to the computational domain $\Omega$ 
(see Equation~\ref{waveequationH})
of the FEM simulations. 

We discretize the computational domain using a 3D mesh generator~\cite{schoberl_netgen}.
This leads to a coarse unstructured tetrahedral mesh containing around $10^3$
tetrahedra. 
During the execution of the simulation this mesh is refined to a finer 
mesh (several refinement steps, in each of which every tetrahedron is refined to eight 
smaller tetrahedra).
Figure~\ref{mesh} shows elements of a refined mesh.
Obviously, by using such unstructured meshes, irregular geometries with nearly 
arbitrary shapes of the scatterers can be resolved ideally.
In order to realize transparent boundary conditions with the PML method we discretize 
the exterior space with prism elements above and below the sample. 
The coarse grid in this case contains 1680 prisms which support second-order finite elements.

In the investigated regime, we assume a dependency of the permittivity of gold on the 
light frequency given by the Drude model:
\begin{equation}
\epsilon(\omega)=1-\frac{\omega_p^2}{\omega(\omega+i \omega_c)}\quad ,
\end{equation}
with the plasma frequency $\omega_p$ and the collisional frequency $\omega_c$.
The used parameters as well as the assumed permittivities of the other present 
materials are given in Table~\ref{srr_para_t5}.

\section{Numerical results}
The discrete problem corresponding to 
Equations~(\ref{waveequationH}) -- (\ref{bloch}) with the parameters of 
Chapter~\ref{parameterchapter} leads to a matrix equation with 
$N=28\,600$\,unknowns for the coarse grid, resp. $N=133\,326$\,unknowns
for the one time uniformly refined grid.
This equation is solved by LU factorization 
(package 'PARDISO'~\cite{PARDISO}) on a standard 64bit PC
({\it AMD Opteron}).
Typical computation times are 30\,sec ($N=28\,600$), resp.~5\,min ($N=133\,326$).

\begin{figure}[hbt]
\centering
\psfrag{(a)}{(a)}
\psfrag{(b)}{(b)}
\includegraphics[width=0.75\textwidth]{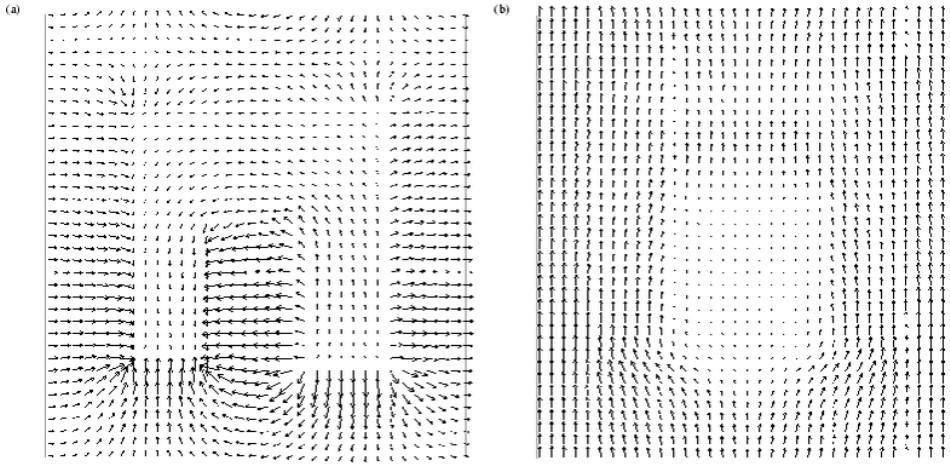}
\caption{Cross sections through the projection onto the $x-y-$plane of 
the 3D vectorial solutions at the center of the SRR ($15\,$nm above the ITO layer).
(a) Electric field, (b) magnetic field. Parameters: $\lambda=1560\,$nm, incident wave
polarized in the $x-z-$plane ($\Field{E}$), resp. in the $y-z-$plane ($\Field{H}$), perpendicular
incidence. Please note that the magnetic field in the inner of the SRR points mainly in 
the $z-$direction, which is not seen in this presentation of the data.{\footnotesize (See original publication for images with higher resolution.)}
}
\label{solution_0deg}
\end{figure}

Cross-section of the 3D vectorial solutions, 
$\Field{E}(\vec{x})$ and $\Field{H}(\vec{x})$, 
for $z=const.$ at the center of 
the SRR is shown in Figure~\ref{solution_0deg}. 
In this case the wavelength of the incident plane wave is $\lambda=1.56\,\mu$m, 
and the incidence is perpendicular ($\vec{k}=|2\pi/\lambda|\times(0,\,0,\,-1)$).
The polarization of the electric field of the incident wave is in the $x-z-$plane, 
i.e., $E_0=(1.0,\,0,\,0)\,$V/m.

In order to obtain the transmission coefficient 
for light scattered into the zero$^th$ diffraction order
we perform a Fourier transform 
of the solution at the bottom of the computational domain ($z=z_{sub}$):

\begin{equation}
A_i(\vec{k}_{FC})=
\frac{1}{a_x\cdot a_y}
\int_{-a_x/2}^{a_x/2}
\int_{-a_y/2}^{a_y/2}
E_i(x,y,z_{sub})
\exp(-i \vec{k}_{FC}\vec{x} ) 
dx\, dy
\end{equation}

where $\vec{k}_{FC}$ is the projection of the wavevector of the zero$^th$ diffraction order
onto the  $x-y-$plane ($\vec{k}_{FC}=0$ for perpendicular incidence). 
In accordance with the convention used in the experiments~\cite{Linden2004a,Enkrich2005a}
we then define the transmission as $T=\frac{I_t}{\alpha I_{in}}$, where $\alpha$ corresponds to 
the transmission of a plane wave through a sample without SRR's, $I_{in}$ is the intensity
of the incoming wave, and $I_t$ is the intensity of the transmitted zero-order plane 
wave corresponding to the Fourier coefficients $A_i$.
The reflection, $R$, is defined accordingly. 

\begin{figure}[hbt]
\centering
\includegraphics[width=0.5\textwidth]{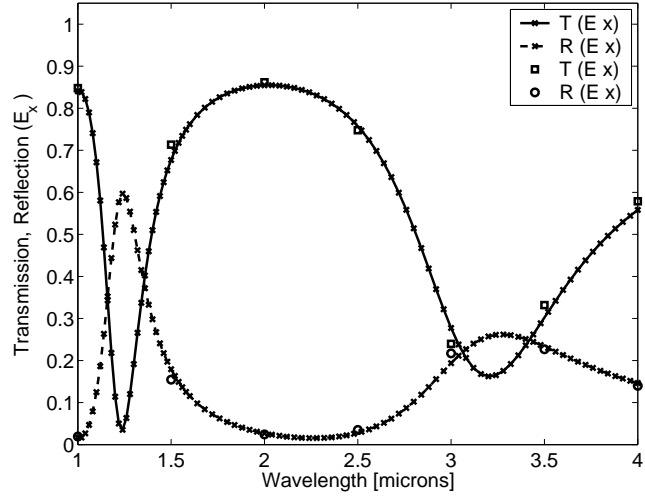}
\caption{Transmission and reflection of a lightfield incident onto an SRR. 
Parameters correspond to the experiment of Linden et al.~\cite{Linden2004a}, 
Figure 2A. Please note the very good quantitative agreement with the 
experimental results. 
Transmission and reflection have been simulated on a coarse grid (x's and solid, 
resp.~dashed, lines) and on a refined grid (squares/circles).{\footnotesize (See original publication for images with higher resolution.)}
}
\label{convergence_srr1}
\end{figure}

Figure~\ref{convergence_srr1} shows the calculated transmission and reflection of 
a light field with perpendicular incidence on a SRR. 
The comparison of the results obtained from simulations 
on the coarse grid (with $N=76\,014$ in this case) with 
the results obtained on a refined grid ($N=374\,366$) for several wavelengths 
(squares and circles in Figure~\ref{convergence_srr1})
shows that for the investigated regime simulations with around $10^5$ unknowns
are already well converged and the error in $T$ and $R$ can be estimated to be 
less than few percent.

\begin{figure}[hbt]
\centering
\includegraphics[width=0.5\textwidth]{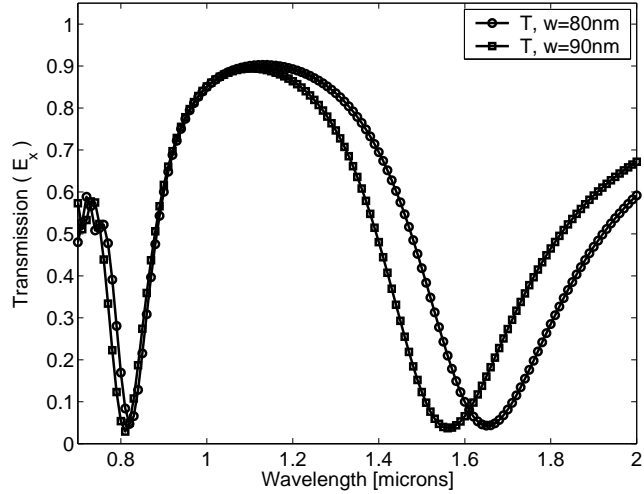}
\caption{Transmission spectra of light fields incident onto an SRR. 
All SRR parameters correspond to Table~\ref{srr_para_t5} except for $w$
which is $w=80\,$nm (circles), resp.~$w=90\,$nm (squares).
{\footnotesize (See original publication for images with higher resolution.)}
}
\label{spectra_w}
\end{figure}

The physics of the resonances of the SRR excited by the incident light field
(around $\lambda=1.25\,\mu$m and  $\lambda=3.2\,\mu$m in Figure~\ref{convergence_srr1})
has been explained in detail in previous works~\cite{Linden2004a,Enkrich2005a}.
As an example of the strong dependence of the resonances on the geometrical 
parameters of the SRR's we present in Figure~\ref{spectra_w} two transmission 
spectra for light fields (perpendicular incidence, $\Field{E}-$field in the 
$x-z-$plane) incident on SRR's with parameters according to Table~\ref{srr_para_t5}, except 
for the width of the lower bar of the 'U' which is varied by a width of 10\,nm.

\begin{figure}[hbt]
\centering
\psfrag{(a)}{(a)}
\psfrag{(b)}{(b)}
\includegraphics[width=1.0\textwidth]{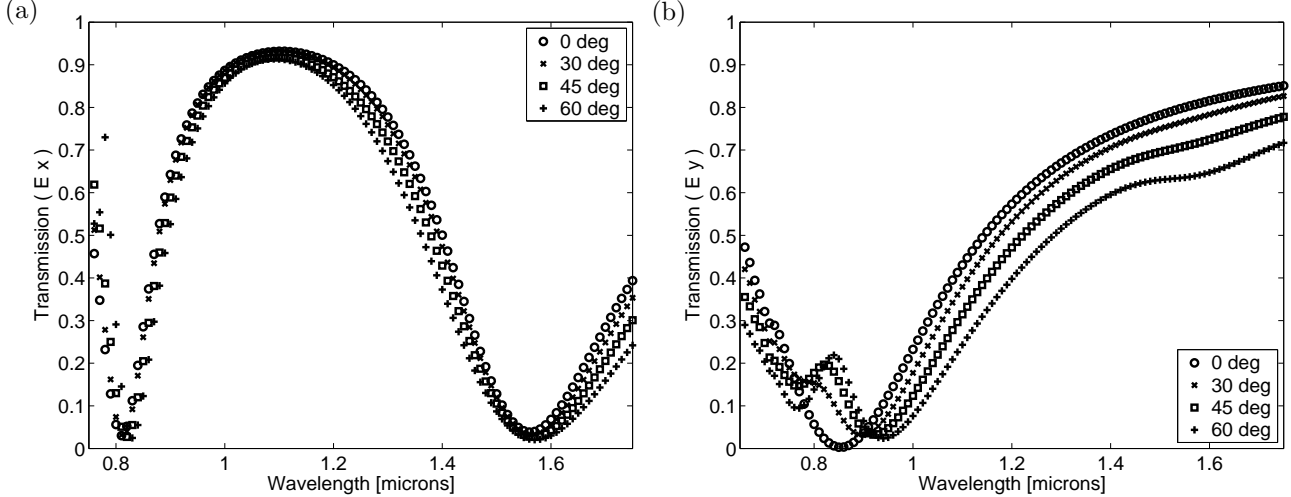}
\caption{Transmission spectra of light fields incident onto an SRR for different 
angles of incidence. 
Polarization of the incident $\Field{E}-$field in $x-$direction (a), 
resp.~$y-$direction (b).
{\footnotesize (See original publication for images with higher resolution.)}
}
\label{spectra_alpha}
\end{figure}

\begin{figure}[hbt]
\centering
\psfrag{x}{$x$}
\psfrag{y}{$y$}
\psfrag{(a)}{$\omega t=0$}
\psfrag{(b)}{$\omega t=0.25\,\pi$}
\psfrag{(c)}{$\omega t=0.5\,\pi$}
\psfrag{(d)}{$\omega t=0.75\,\pi$}
\psfrag{(e)}{$\omega t=1.0\,\pi$}
\psfrag{(f)}{$\omega t=1.25\,\pi$}
\psfrag{(g)}{$\omega t=1.5\,\pi$}
\psfrag{(h)}{$\omega t=1.75\,\pi$}
\includegraphics[width=0.75\textwidth]{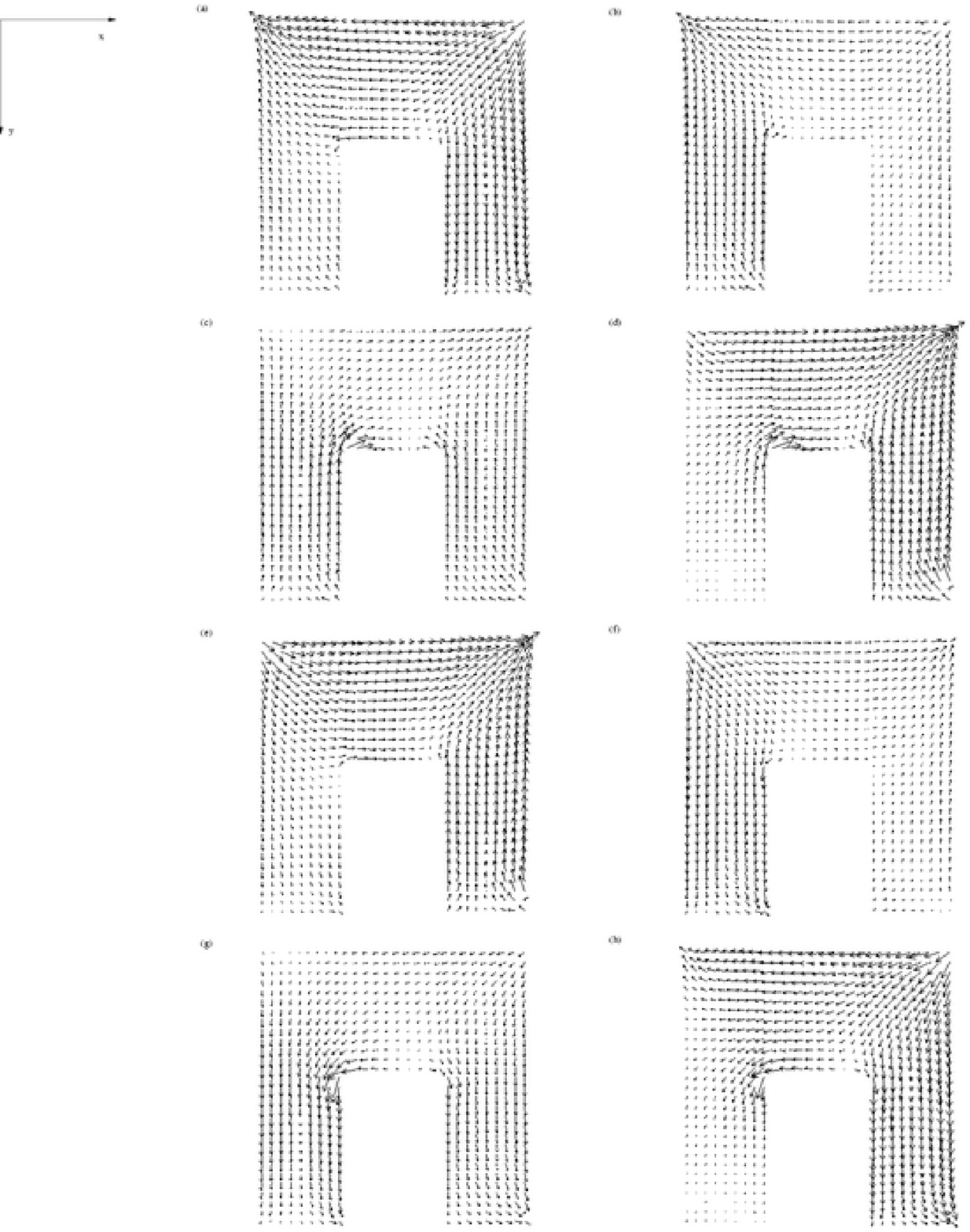}
\caption{Cross sections of the solution in $x-y-$direction in the region of the SRR and 
at the center of the SRR ($15\,$nm above the ITO layer).
Shown is the projection of the real part of the 
$\Field{E}-$field onto the $x-y-$plane for different
phases $\omega t$ of the time-harmonic solution.
The exciting light field has a wavelength of $\lambda=780\,$nm and is 
incident with an angle of $\alpha=60\,$deg (solution corresponds to the low-wavelength
resonance  in  Fig.~\ref{spectra_alpha}b).
{\footnotesize (See original publication for images with higher resolution.)}
}
\label{solution_60deg_phases}
\end{figure}

By applying light fields which are incident onto the arrays of SRR under oblique
angle, it is possible to observe several effects:
Resonances can be excited now by the magnetic field component of the incident 
light field, resonances are shifted, and new, asymmetric resonances can be observed.
Figure~\ref{spectra_alpha} shows simulated spectra for different angles of incidence. 
In Fig.~\ref{spectra_alpha}a, the $\vec{k}$ vector of the incident plane wave 
is in the $y-z-$plane and encloses angles of $\alpha=0,30,45,60\,$deg with 
the $z-$axis, 
in Fig.~\ref{spectra_alpha}b, the $\vec{k}$ vector of the incident plane wave 
is in the $x-z-$plane and again encloses angles of $\alpha=0,30,45,60\,$deg with 
the $z-$axis.
The electric field is polarized in $x-$direction (a), resp.~$y-$direction (b)
 (compare Figure 2 in ref~\cite{Enkrich2005a}).
All simulations are performed using a refined grid with $N=133\,326$\,unknowns.
It is again noted that in the case of oblique incidence special care has to be 
taken about the appropriate periodic boundary conditions in $x-$ and 
$y-$direction (see Equation~\ref{bloch}).
In very good quantitative agreement with the experimental results~\cite{Enkrich2005a}
the main effect of an increased angle $\alpha$ in case (a) seems to be an 
increase in linewidth of the resonance around 1.55\,$\mu$m.
This can intuitively explained by the fact that in this case the resonance is 
excited by a coupling of the electric field to the lower bar of the `U'-shaped 
SRR, which is not affected by a change of angle $\alpha$. However, the broadening
can probably be attributed to the fact that the cross-section of the lower bar has
a substantially larger horizontal dimension than vertical dimension.

The resonance around 1.55\,$\mu$m can not be excited by the electric field in case (b).
But, as the angle  $\alpha$ is increased the $z-$component of the 
magnetic field of the incident wave is increased, too. This component couples
to the resonance of the SRR. 
Therefore, around $\lambda=1.55\,\mu$m in case (b), and for $\alpha=60\,$deg 
a weak resonance of the SRR can be observed. Again, position as well as strength 
of the resonance are in quantitative very good agreement with the 
experimental results~\cite{Enkrich2005a}.

Interestingly, in case (b) the resonance around $\lambda=850\,$nm is shifted to higher 
wavelengths for increased angle  $\alpha$, which can be attributed to the phase shift 
between the induced currents in the two upper arms of the 'U'-shaped SRR.
Moreover, another resonance around $\lambda=780\,nm$ appears which is not excited in 
the case of perpendicular incidence of the plane wave.
Figure~\ref{solution_60deg_phases} shows cross-sections through the 
real part of the electric field 
present in the SRR at different phases of the time-harmonic solution.
From these distributions it can be seen that the resonance consists in oscillations
of the electric field in both upper arms of the 'U', with different amplitudes and 
phases, and an oscillation in the lower bar of the 'U', which leads to an accumulation 
of charge at the outer bottom edges.

\section{Conclusion}
In this paper we have  
investigated rigorous numerical solutions of the 3D electromagnetic 
scattering problem of plane 
waves incident onto a periodic array of split-ring resonators in the 
optical regime.
The solutions, obtained on
standard personal computers, show an excellent agreement with experimental 
observations.

Future research directions include the investigation of systems with 
a macroscopically negative refractive index and nonlinear properties of 
metamaterials.
\acknowledgements
We acknowledge support by  
the priority programme SPP 1113 of the Deutsche Forschungsgemeinschaft, 
DFG, and by the German Federal Ministry of
Education and Research, BMBF, under contract No.~13N8252 ({\sc HiPhoCs}).

\bibliography{/home/numerik/bzfburge/texte/biblios/phcbibli,/home/numerik/bzfburge/texte/biblios/group05}   

\begin{thebibliography}{10}

\bibitem{Linden2004a}
S.~Linden, C.~Enkrich, M.~Wegener, C.~Zhou, T.~Koschny, and C.~Soukoulis,
  ``Magnetic response of metamaterials at 100 {T}erahertz,'' {\em Science} {\bf
  306}, p.~1351, 2004.

\bibitem{Enkrich2005a}
C.~Enkrich, M.~Wegener, S.~Linden, S.~Burger, L.~Zschiedrich, F.~Schmidt,
  C.~Zhou, T.~Koschny, and C.~M. Soukoulis, ``Magnetic metamaterials at
  telecommunication and visible frequencies.''
\newblock {\it Phys. Rev. Lett., in press,} preprint available from
  http://arxiv.org/pdf/cond-mat/0504774, 2005.

\bibitem{Zschiedrich2005b}
L.~Zschiedrich, R.~Klose, A.~Sch\"adle, and F.~Schmidt, ``A new finite element
  realization of the perfectly matched layer method for {H}elmholtz scattering
  problems on polygonal domains in 2{D},'' {\em J. Comp. Phys., {\it in press}}
  , 2005.

\bibitem{Burger2005a}
S.~Burger, R.~Klose, A.~Sch\"adle, F.~Schmidt, and L.~Zschiedrich, ``{FEM}
  modelling of 3d photonic crystals and photonic crystal waveguides,'' in {\em
  Integrated Optics: Devices, Materials, and Technologies IX},  Y.~Sidorin and
  C.~A. W\"achter, eds.,  {\bf 5728}, pp.~164--173, Proc. SPIE, 2005.

\bibitem{Veselago1968a}
V.~G. Veselago, ``The electrodynamics of substances with simultaneously
  negative values of $\epsilon$ and $\mu$,'' {\em Sov. Phys. Usp.} {\bf 10},
  p.~509, 1968.

\bibitem{Smith2000a}
D.~R. Smith, W.~J. Padilla, D.~C. Vier, S.~C. Nemat-Nasser, and S.~Schultz,
  ``Composite medium with simultaneously negative permeability and
  permittivity,'' {\em Phys. Rev. Lett.} {\bf 84}, p.~4184, 2000.

\bibitem{Shelby2001a}
R.~A. Shelby, D.~R. Smith, and S.~Schultz, ``Experimental verification of a
  negative index of refraction,'' {\em Science} {\bf 292}, p.~77, 2001.

\bibitem{Pendry2000a}
J.~B. Pendry, ``Negative refraction makes a perfect lens,'' {\em Phys. Rev.
  Lett.} {\bf 85}, p.~3966, 2000.

\bibitem{Sakoda2001a}
K.~Sakoda, {\em Optical Properties of Photonic Crystals}, Springer-Verlag,
  Berlin, 2001.

\bibitem{Hohage03a}
T.~Hohage, F.~Schmidt, and L.~Zschiedrich, ``Solving {T}ime-{H}armonic
  {S}cattering {P}roblems {B}ased on the {P}ole {C}ondition {I}: {T}heory,''
  {\em SIAM J. Math. Anal.} {\bf 35}(1), pp.~183--210, 2003.

\bibitem{Hohage03b}
T.~Hohage, F.~Schmidt, and L.~Zschiedrich, ``Solving {T}ime-{H}armonic
  {S}cattering {P}roblems {B}ased on the {P}ole {C}ondition {II}: {C}onvergence
  of the {PML} {M}ethod,'' {\em SIAM J. Math. Anal.} {\bf 35}(3), pp.~547--560,
  2003.

\bibitem{Zschiedrich2005a}
L.~Zschiedrich, S.~Burger, R.~Klose, A.~Sch\"adle, and F.~Schmidt, ``{JCMmode}:
  {An} adaptive finite element solver for the computation of leaky modes,'' in
  {\em Integrated Optics: Devices, Materials, and Technologies IX},  Y.~Sidorin
  and C.~A. W\"achter, eds.,  {\bf 5728}, pp.~192--202, Proc. SPIE, 2005.

\bibitem{Deuflhard2003a}
P.~Deuflhard, F.~Schmidt, T.~Friese, and L.~Zschiedrich, {\em Adaptive
  Multigrid Methods for the Vectorial Maxwell Eigenvalue Problem for Optical
  Waveguide Design}, pp.~279--293.
\newblock Mathematics - Key Technology for the Future, Springer-Verlag, Berlin,
  2003.

\bibitem{Monk2003a}
P.~Monk, {\em Finite Element Methods for {M}axwell's Equations}, Claredon
  Press, Oxford, 2003.

\bibitem{schoberl_netgen}
J.~Sch\"oberl, ``Netgen.''
\newblock Available from http://www.hpfem.jku.at.

\bibitem{PARDISO}
{O. Schenk {\it et al.}}, ``Parallel sparse direct linear solver {PARDISO}.''
\newblock Department of Computer Science, Universit\"at Basel.

\end{thebibliography}
\bibliographystyle{spiebib}   

\end{document}